\documentclass{llncs}

\usepackage{inputenc}
\usepackage{algorithm}
\usepackage{float}
\usepackage{units}
\usepackage{graphicx}
\usepackage{multirow}
\usepackage{extarrows}
\usepackage{ulem}
\usepackage{balance}
\usepackage{cite}
\usepackage[colorlinks]{hyperref}
\usepackage{breakurl}
\usepackage{microtype}
\usepackage{times}
\usepackage{epsfig}
\usepackage[TABBOTCAP]{subfigure}
\usepackage{tabularx}
\usepackage{color}
\usepackage{xspace}
\usepackage{thumbpdf}
\usepackage{listings}
\usepackage{verbatim}
\usepackage{booktabs}
\usepackage{colortbl}
\usepackage[font=small,labelfont=bf]{caption}
\newcommand{\cref}[1]{Chapter~\ref{#1}}

\newcommand{\first}{\emph{(i)}~}
\newcommand{\second}{\emph{(ii)}~}
\newcommand{\third}{\emph{(iii)}~}
\newcommand{\fourth}{\emph{(iv)}~}

\newcommand{\ie}{i.e., \@}

\newcommand{\cf}{cf. \@}

\definecolor{darkgreen}{rgb}{0,0.5,0}
\definecolor{brown}{rgb}{0.7,0.3,0}
\definecolor{darkblue}{rgb}{0,0,0.5}

\newcounter{fn1}
\newcounter{fn2}
\newcounter{fn3}
\newcounter{fn4}
\newcounter{fn5}
\begin{document}
\title{\texttt{traIXroute}: Detecting IXPs in \texttt{traceroute} paths}
\author{George Nomikos \and Xenofontas Dimitropoulos}
\institute{Foundation of Research and Technology Hellas (FORTH), Greece 
\email{\{gnomikos,fontas\}@ics.forth.gr}
}

\maketitle

\begin{abstract} 
Internet eXchange Points (IXP) are critical components of the Internet infrastructure that affect its performance, evolution, security and economics. In this work, we introduce techniques to augment the well-known \texttt{traceroute} tool with the capability of identifying if and where exactly IXPs are crossed in end-to-end paths. Knowing this information can help end-users have more transparency over how their traffic flows in the Internet. Our tool, called \texttt{traIXroute}, exploits data from the PeeringDB (PDB) and the Packet Clearing House (PCH) about IXP IP addresses of BGP routers, IXP members, and IXP prefixes. We show that the used data are both rich, \ie we find 12,716 IP addresses of BGP routers in 460 IXPs, and mostly accurate, \ie our validation shows 92-93\% accuracy. In addition, 78.2\% of the detected IXPs in our data are based on multiple diverse evidence and therefore help have higher confidence on the detected IXPs than when relying solely on IXP prefixes. To demonstrate the utility of our tool, we use it to show that one out of five paths in our data cross an IXP and that paths do not normally cross more than a single IXP, as it is expected based on the valley-free model about Internet policies. Furthermore, although the top IXPs both in terms of paths and members are located in Europe, US IXPs attract many more paths than their number of members indicates.
\end{abstract}
	
\section{Introduction\label{introduction}}
A few hundred IXPs worldwide host more than one hundred thousand interconnections between Autonomous Systems (ASes)~\cite{ager2012anatomy},~\cite{Chatzis:2013},~\cite{Giotsas:2013}. As critical components of the Internet infrastructure, IXPs influence its expansion~\cite{dhamdhere2010internet}, performance~\cite{Ahmad:2010}, and security~\cite{spamhaus}. However, their centralized nature is also a limitation that can be exploited for mass surveillance of Internet users or for targeted attacks. Although IXPs exist since the early days of the Internet, they have recently attracted intense interest from the academic community in part because the last decade the Internet topology is flattening~\cite{dhamdhere2010internet},~\cite{gill2008flattening},~\cite{gregori2011impact},~\cite{labovitz2011internet}, which implies an even more central role for IXPs.

In this work we extend the well-known and widely-used traceroute tool with the capability of inferring if and where an IXP was crossed. This is useful not only for end-users in having more transparency over where their traffic goes, but also for operators in troubleshooting end-to-end paths and for researchers in understanding the evolving IXP ecosystem. Our tool, called \texttt{traIXroute}, detects IXPs based on data from the PeeringDB (PDB) and the Packet Clearing House (PCH). In particular, it uses the \first exact IP addresses of BGP routers connected to IXP subnets; \second IXP member ASes; \third IXP prefixes; and \fourth IP addresses to AS mappings; and combines multiple information to detect IXPs with higher confidence than simply relying on IXP prefixes.

Our second contribution is that we evaluate the coverage and accuracy of the IXP router IP addresses, which we denote with a triplet \{\textit{IP address} $\longrightarrow$ \textit{IXP}, \textit{AS}\}, in PDB and PCH. We find in total 12,716 triplets for 460 IXPs worldwide. Using the exact router IXP addresses along with checking the IXP membership of the two adjacent ASes, we classify 78.2\% of the IXP paths. Therefore, in most cases we can detect an IXP with strong evidence. In addition, we find that 92-93\% of the triplets \{\textit{IP address} $\longrightarrow$ \textit{IXP}, \textit{AS}\} extracted from PDB and PCH are consistent with the corresponding information extracted from live BGP sessions of route collectors at IXPs. 

Third, to illustrate how \texttt{traIXroute} can be useful, in particular for researchers in Internet measurement studies, we use it to answer the following questions: \first how often paths cross IXPs? \second which IXPs attract most paths? and \third how many IXPs are encountered per path? We apply \texttt{traIXroute} on 31.8 million traceroute probes collected from the ark measurement infrastructure~\cite{ark}. We find that approximately one out of five paths crossed an IXP and that IXP-paths normally cross no more than a single IXP. The IXP hop is located on average near the 6th hop at the middle of the route. Finally, we show that the top IXPs in terms of paths differ in part from the top IXPs in terms of AS members.

The rest of this paper is structured as follows. In the next section, we discuss the related work and provide background into the problem of detecting IXPs in traceroute paths. Next, in Section~\ref{tool} we describe \texttt{traIXroute} and its IXP detection techniques. In Section~\ref{eval}, we evaluate the coverage and accuracy of the data used by \texttt{traIXroute} and discuss the hit rate of its detection rules. Finally, in Section~\ref{usecase} we outline our IXP measurement study using \texttt{traIXroute} and in Section~\ref{conclusion} we conclude.

\section{Related Work and Background}\label{relatedwork}
Previous studies have examined the problem of mapping traceroute paths to AS-level paths~\cite{Chen:2009},~\cite{mao2003towards}. Mapping IP addresses to ASes is not straightforward because routers can reply with source IP addresses numbered from a third-party AS. These studies ignore hops with IXP IP addresses. These addresses are used to number BGP router interfaces connected to the IXP subnet and it is hard to identify to which AS they belong. 

Besides, a group of previous studies, starting with Xu~{\it et al}.~\cite{xu2004properties} and then followed by He~{\it et al}.~\cite{he2009lord} and Augustin~{\it et al}.~\cite{augustin2009ixps}, focus on inferring participating ASes and peerings at IXPs from {\it targeted} traceroute measurements. Compared to these studies, our goal is different: we build a general-purpose traceroute tool, while they aim at discovering as many peering links as possible. The basic methodology developed in~\cite{xu2004properties} and then significantly extended in~\cite{he2009lord} and~\cite{augustin2009ixps} detects IXPs based on assigned IP address prefixes and uses various heuristics to infer peering ASes.  The seminal work of Augustin~{\it et al}.~\cite{augustin2009ixps} exploited also data for BGP routers at IXP, but by querying 1.1K BGP Looking Glass servers, which had significant processing cost. In contrast, we extract corresponding data from PDB and PCH, with low processing cost, and show that they are both rich and mostly accurate. 

Recently, Giotsas~{\it et al}.~\cite{giotsasmapping} introduced techniques to identify the physical facility where ASes interconnect using targeted traceroute measurements and a combination of publicly available facility and IXP based information.

\begin{figure}
	\centering
	\includegraphics[scale=0.25]{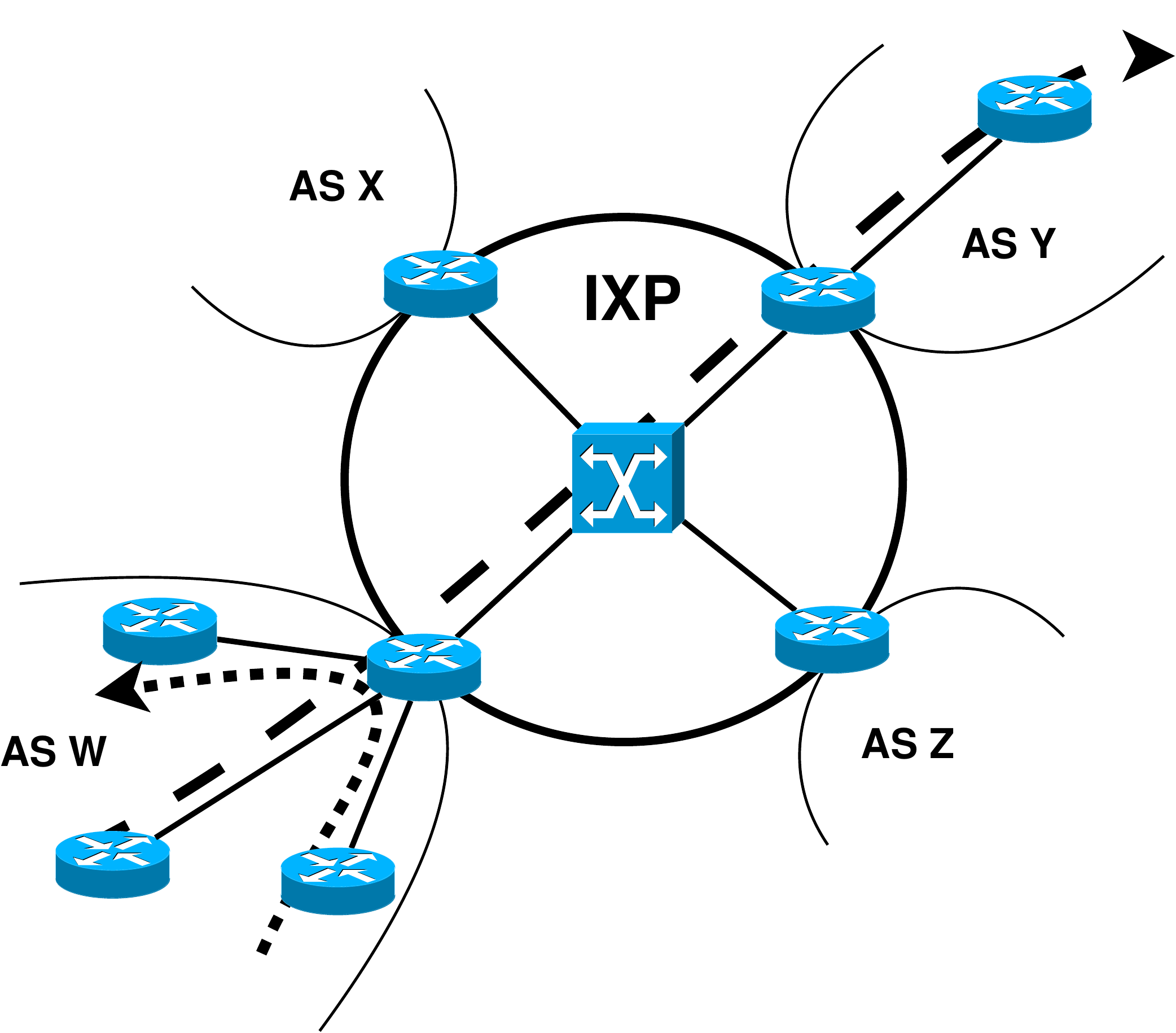}
	\hspace{0.5cm}
	\caption{Example IXP connected to four ASes. The dotted traceroute path could include a reply with an IXP IP address, even if the IXP is not crossed. Our goal is to identify paths that cross the IXP, like the dashed one.}
	\label{fig:challenges}
\end{figure}

Our starting point in this work is that {\bf observing an IP address from an IXP prefix is not sufficient evidence to conclude that the IXP was crossed}. This happens for multiple reasons: \first the available IXP IP address prefix data may be inaccurate; \second IXPs could use allocated addresses not only in the IXP subnet but also in other operational subnets; and \third third-party IP addresses from IXP subnets. To illustrate the latter consider the following example (cf. Figure~\ref{fig:challenges}). A router connected to the IXP fabric could reply to traceroute probes using a source IP address from any of its interfaces, including the interface on the IXP subnet. Traceroute paths that do not cross the IXP, like the dotted one in Figure~\ref{fig:challenges}, can include a reply with a source IP address from the IXP subnet. Therefore, the path appears to have an IP address from an IXP subnet, even if the IXP is not crossed. Our goal is to detect paths that cross the IXP fabric, like the dashed one in Figure~\ref{fig:challenges}.

To be more confident that an IXP is crossed, we exploit specific information about the IP addresses of BGP router interfaces connected to the IXP subnet. This data enable us also to associate IP addresses to ASes and IXPs. Furthermore, we check if the ASes before and after the IXP IP address are members of the candidate IXP based on the IXP membership data from PCH and PDB, which have not been explored in the previous studies for this purpose. 

\section{\textsc{traIXroute} Design and Heuristics}\label{tool}
In this section, we first outline the design of \texttt{traIXroute} and then its IXP detection heuristics. 

\subsection{\textsc{traIXroute} Design}
\begin{figure}
	\centering
	\resizebox{340pt}{140pt}{
		\includegraphics{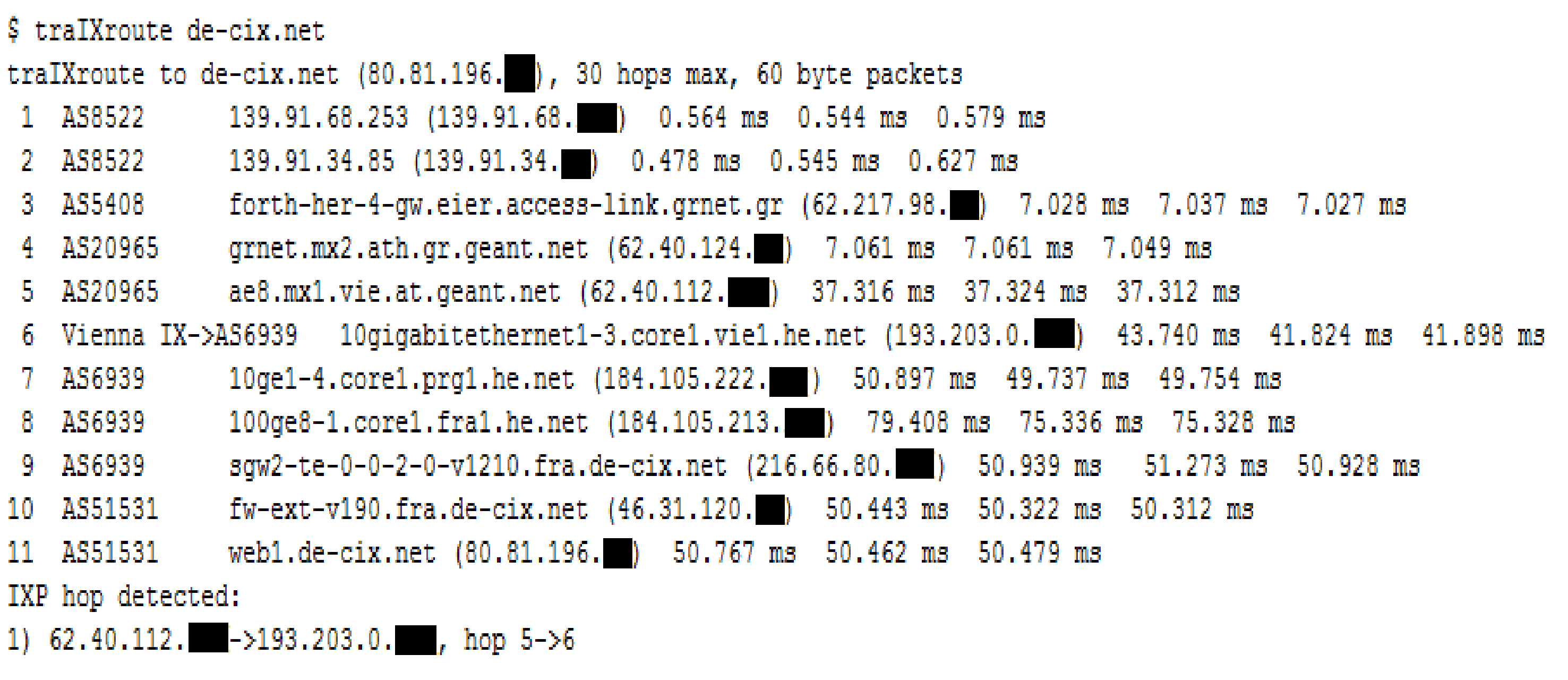}
	}
	\caption{Example output of \texttt{traIXroute}.}
	\label{fig:traceroute_output}
\end{figure}

\texttt{traIXroute} is written in python and operates like \texttt{traceroute}. It can be configured to use either the standard \texttt{traceroute} tool in the background or the \texttt{scamper} tool \cite{luckie2010scamper}, which implements the Paris traceroute technique~\cite{augustin2007multipath}. It has a modular design and can be easily extended with new IXP data and detection rules. An example of the output of \texttt{traIXroute} is shown in Figure~\ref{fig:traceroute_output}. In this example the Vienna IX is detected between hops 5 and 6. The tool also prints the AS that corresponds to each hop based on simple origin AS lookups. \texttt{traIXroute} exploits three datasets to identify IXPs in traceroute paths which can be updated automatically from the command line:

\begin{enumerate}
	\item \textbf{IXP Memberships Dataset:} We use IXP membership data from the \textsc{PeeringDB} (PDB)\cite{pdb} and the \textsc{Packet Clearing House} (PCH)\cite{pch}. They provide: 1) exact IP addresses of router interfaces connected to the IXP network; and 2) the ASes  which these routers belong to. Therefore, this dataset provides an association from IXP IP addresses to ASes and IXPs, i.e., a triplet of the form \{\textit{IP address} $\longrightarrow$ \textit{IXP}, \textit{AS}\}, which we mainly exploit in our heuristics.
	
	\item \textbf{IXP IP Address Prefixes Dataset:} We use, in addition, two datasets of IPv4 address prefixes assigned to IXPs. The first is provided by PDB, while we extract the second from PCH. These addresses are typically used to number the interfaces of the BGP routers connected to the IXP subnet. We organize the dataset in the form \{\textit{IP prefix} $\longrightarrow$ \textit{IXP}\} to map IP addresses to IXPs.
	
	\item \textbf{Routeviews Prefix to AS mappings Dataset:} We use IP address prefix to AS mappings, i.e., \{\textit{IP prefix} $\longrightarrow$ \textit{AS}\}, provided by CAIDA~\cite{routeviews_prefixes} based on data from RouteViews~\cite{routeviews}, to associate IP addresses to ASes. Also, we filter the IANA reserved IP addresses, which should not be announced to BGP, to protect from route leaks and other misconfigurations. When encountering multi-origin-as~\cite{moas} IP addresses, we check the IXP membership of all the ASes.
\end{enumerate}

PCH and PDB do not use consistent identifiers for IXPs and therefore if one naively matched the IXP identifiers would introduce artifacts. For this reason, we merge the two datasets by matching the IXP IP addresses, prefixes and names. We ignore matched records that include inconsistent attributes. In addition, we filter data for IXPs marked as inactive. 

\subsection{IXP detection}
Next, we describe our methodology to detect and identify at which hop we cross an IXP in traceroute paths. When observing an IP address from an IXP subnet, we ask what information we know, based on our data, for this and the adjacent IP addresses. In particular, to infer an IXP crossing we follow three steps: 

{\bf (Step 1) - Does the IP address match an exact BGP router IP address from an IXP subnet?} In this case, we have a specific triplet \{\textit{IP address} $\longrightarrow$ \textit{IXP}, \textit{AS}\}, which gives us also additional information about the AS of the router on the IXP. If an exact router IP address is not matched, then we check if an IXP prefix is matched, like in previous works \cite{augustin2009ixps}\cite{he2009lord}. However, in this case we do not have any information about the AS that owns the router. If an IP address in the $k$-th hop of a traceroute path~$IP_k$ belongs to the interface of a router connected to the IXP subnet, then we denote this with~$IP_{k}\xlongrightarrow{inf} IXP,AS_{k}$, where~$IXP$ is the IXP and~$AS_{k}$ the AS of the router. Otherwise, if we can associate~$IP_{k}$ only with an IXP IP prefix, then we denote this with~$IP_{k}\xlongrightarrow{prf} IXP$.

{\bf (Step 2) - Are the adjacent ASes members of the IXP?} We map the IP addresses 1-hop adjacent to the observed IXP IP address to ASes and, considering also the AS of the IXP IP address (if this information is available), we check the IXP membership of the ASes. We distinguish four possible cases: \first both ASes are members, \second-\third only the AS in the left or right of the IXP IP address is a member; and \fourth none of the ASes is an IXP member. Our assessment is based on the available data about the ASes from triplets and from mapping IP addresses to ASes using the Routeviews Prefix to AS mappings Dataset. Such mappings could be wrong~\cite{mao2003towards}, therefore we do not consider this evidence alone conclusive. In addition, if $AS_{k}$ is a member of the $IXP$ based on IXP membership data then we denote this with~$AS_{k} \in IXP$. 

{\bf (Step 3) - Is the IXP link crossed before or after the IXP IP address?} We check this when sufficient information about the ASes is available. 

Our heuristics are applied on a traceroute path in a sliding window fashion, where the length of the window is three. By carefully reasoning about all possible combinations of evidence from Steps 1 and 2 that exist for three subsequent hops, we formulated 16 cases. Each case corresponds to a detection rule. For brevity, we next discuss only the cases (8 in total) that appeared with frequency higher than 1\% in the matched IXP paths. The remaining cases are still supported in \texttt{traIXroute}. In Table~\ref{tab:cases1} we show our detection rules for the most typical scenario, when we observe a single IXP IP address between two non-IXP IP addresses. We also consider the special case, shown in Table~\ref{tab:cases2}, when we observe two adjacent IP addresses from an IXP subnet. In most cases, we can deduce the exact link where the IXP was crossed, which we denote in Tables~\ref{tab:cases1} and~\ref{tab:cases2} as~$a$ or~$b$. We split the rules into \textit{strong} and \textit{weak} evidence rules and order them based on their frequency, as shown in the last column of the tables (\cf Section~\ref{eval}). 

\begin{table*}[t]
	\centering \small
	%\advance\leftskip-0.6cm
	\resizebox{340pt}{90pt}{
		\begin{tabular}{|c|c|c|c|c||c|}
			\hline
			\multicolumn{6}{|c|}{\textbf{One IXP IP Address between two non-IXP IP addresses}}\\ \hline\hline
			\textit{Rules} & \multicolumn{3}{c|}{\textit{Hop Window}}  & \textit{Assessment} & \textit{Hit Rate} \\ \hline
			&\multicolumn{3}{c|}{\centering \includegraphics[scale=0.35]{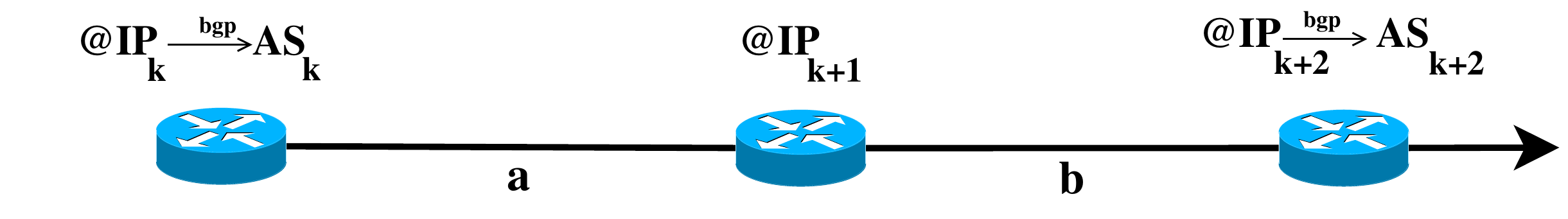}}& & \\\hline \hline
			
			1.1 & ~~~\shortstack{$AS_{k} \in  IXP$}~~~~~ 
			& ~~~\shortstack{$IP_{k+1}\xlongrightarrow{inf} IXP,AS_{k+1}$~~~\\ $AS_{k+1} = AS_{k+2} \neq AS_{k}$} 
			&\shortstack{$AS_{k+2}\in  IXP$}  
			& $a\rightarrow IXP$  & 65.57$\%$ \\ \hline
			
			1.2 &\shortstack{$AS_{k}\in  IXP$}
			&\shortstack{$IP_{k+1}\xlongrightarrow{inf} IXP,AS_{k+1}$\\$AS_{k+1} \neq AS_{k} \neq AS_{k+2}$} 
			&\shortstack{$AS_{k+2}\notin  IXP$}  
			&$a\rightarrow IXP$  & 8.79$\%$ \\ \hline
			
			1.3 &\shortstack{$AS_{k}\in  IXP$}
			&\shortstack{$IP_{k+1}\xlongrightarrow{inf} IXP,AS_{k+1}$\\$AS_{k+1} \neq AS_{k} \neq AS_{k+2}$} 
			&\shortstack{$AS_{k+2}\in  IXP$}  
			&$a \text{ or } b\rightarrow IXP$  & 2.5$\%$\\ \hline \hline
			
			1.4 &\shortstack{$AS_{k}\in  IXP$}
			&\shortstack{$IP_{k+1}\xlongrightarrow{prf} IXP$}
			&\shortstack{$AS_{k+2}\notin IXP$}  
			&$a \rightarrow IXP$  & 7.7$\%$ \\ \hline
			
			1.5 &\shortstack{$AS_{k}\notin  IXP$}
			&\shortstack{$IP_{k+1}\xlongrightarrow{prf} IXP$}
			&\shortstack{$AS_{k+2}\in  IXP$} 
			&$b \rightarrow IXP$  & 5.55$\%$\\ \hline
			
			1.6 &\shortstack{$AS_{k}\notin  IXP$}
			&\shortstack{$IP_{k+1}\xlongrightarrow{inf} IXP,AS_{k+1}$\\$AS_{k+1} = AS_{k+2} \neq AS_{k}$}
			&\shortstack{$AS_{k+2}\in IXP$}
			&$b\rightarrow IXP$  & 4.56$\%$ \\ \hline
			
			1.7 &\shortstack{$AS_{k} \notin IXP$}
			&\shortstack{$IP_{k+1}\xlongrightarrow{inf} IXP,AS_{k+1}$\\$AS_{k} \neq AS_{k+1} \neq AS_{k+2}$}
			&\shortstack{$AS_{k+2}\notin IXP$}  
			&$a \text{ or }b \rightarrow IXP$  & 1.21$\%$ \\ \hline
		\end{tabular}}
		\caption{IXP detection rules for a single IXP IP address, based either on IXP interface (inf) or prefix-level (prf) data, between two non-IXP addresses. The rows give the data attributes per hop to check in order to detect an IXP. Rules 1.1 to 1.3 use stronger evidence than Rules 1.4 to 1.7.\label{tab:cases1}}	
	\end{table*}
	
Rules 1.1 to 1.3 match the IP addresses of routers on the IXP subnet, extract information about the adjacent ASes, and find that both ASes are members of the IXP. In the Rules 1.1 and 1.2 the IXP is crossed in the first hop. The Rule 1.2 is otherwise the same with the Rule 1.1, but without information for $AS_{k+2}$. Finally, the Rule 1.3 is also identical otherwise, but with $AS_{k+2} \neq AS_{k+1}$. These three rules check multiple criteria and exploit data about triplets, which give also an association from IP addresses to ASes with high accuracy (\cf Section~\ref{sec:val}). We therefore consider that these rules rely on stronger evidence than the Rules 1.4 to 1.7. 
	
The Rules 1.4 and 1.5 do not match a triplet, but only an IXP prefix. In addition, we find that one of the two adjacent ASes is a member of the IXP. Based on this evidence, we consider that an IXP may have been crossed. However, we have much weaker evidence than when Rules 1.1-1.3 hold. \texttt{traIXroute} marks these cases as potential IXP crossing. Similarly, the Rules 1.6 and 1.7 match an IP address from a triplet, however only one or none of the adjacent ASes is a member of the IXP. We also have weaker evidence in these detections.
	
Finally, the Rule 2 in Table~\ref{tab:cases2} finds two consecutive IP addresses that match triplets from the same IXP. The ASes in the triplets are also found members of the IXP. We consider this also as strong evidence for IXP detection, since multiple evidence indicate so. This is a particularly interesting case, as it indicates that the IXP fabric may have been crossed twice. In other words, we observe in few cases a type of "ping pong" routing over the IXP fabric.
	
	\begin{table}[t]
		\centering
		\resizebox{230pt}{35pt}{
			\begin{tabular}{|c|c|c|c||c|}
				\hline		
				\multicolumn{5}{|c|}{\textbf{Two Consecutive IXP IP Addresses}}\\ \hline\hline
				\textit{Rules} & \multicolumn{2}{c|}{\textit{Hop Window}}  & \textit{Assessment} & \textit{Hit Rate}\\ \hline
				& \multicolumn{2}{c|}{\centering \includegraphics[scale=0.39]{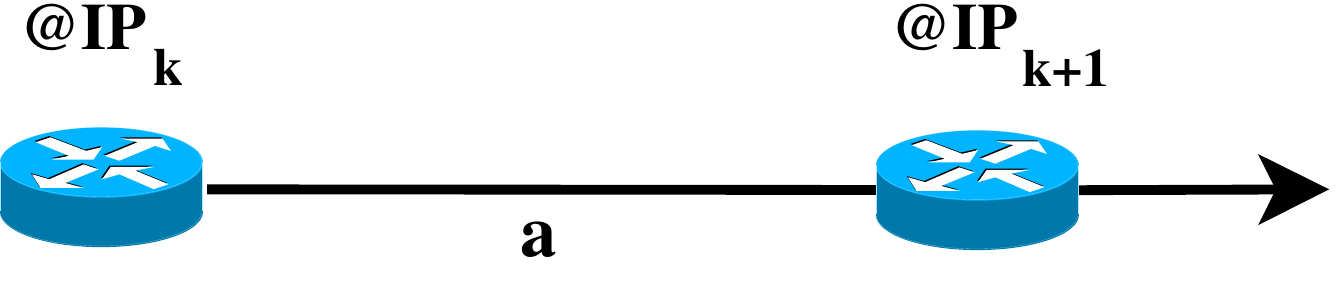}} & &  \\\hline
				
				2.0 &\shortstack{$IP_{k}\xlongrightarrow{inf} IXP,AS_{k}$} 
				&\shortstack{$IP_{k+1}\xlongrightarrow{inf} IXP,AS_{k+1}$\\$AS_{k+1} \neq AS_{k}$} 
				&$a\rightarrow IXP$ & 1.36$\%$ \\ \hline
				
			\end{tabular}}		
			\caption{IXP detection rule for two subsequent IXP IP addresses based on IXP interface (inf) data. The rows give the data attributes per hop which are checked to deduce an IXP.\label{tab:cases2}}
		\end{table}

\section{Evaluation}\label{eval}
In this section, we evaluate and validate our methodology. We downloaded the IXP Memberships Dataset and the IXP IP Address Prefixes Dataset from PDB and PCH on January, the 10th 2015. Our Routeviews Prefix to AS mappings Dataset was downloaded from CAIDA on January, the 20th 2015.

\subsection{Data Coverage and Hit Rates}
PDB includes membership data for 448 (88\%) out of the 509 IXPs in the database. Similarly, PCH provides membership data for 343 (74\%) out of the 466 IXPs it includes. PDB and PCH provide membership data for 100\% and 62\%, accordingly, out of the top-50 IXPs (sorted by the number of their AS members). Besides, 312 of the IXPs in PDB and 343 of the IXPs in PCH provide IXP IP address prefixes. After merging, the combined dataset has 475 address prefixes for 417 IXPs and a total of 12,716 IXP membership triplets \{\textit{IP address} $\longrightarrow$ \textit{IXP}, \textit{AS}\} for 460 IXPs, i.e., an increase of 38.5\% and 3.2\%, correspondingly, with respect to the largest individual dataset. These statistics along with other details are summarized in~Table~\ref{tab:pchpdb}. For comparison, the April 2009 experiment reported by Augustin~{\it et al}.~\cite{augustin2009ixps} found triplets for 119 IXPs by querying 1.1K BGP Looking Glass servers. 

\begin{table}[t]
	\centering
	\resizebox{225pt}{40pt}{
		\begin{tabular}{|c|c|c|}
			\cline{1-3}
			\textbf{Statistics}  & \textbf{PDB} & \textbf{PCH} \\ \hline \hline 
			\multicolumn{1}{|l|}{$\#$ of IXPs} & 509 & 466\\ \hline
			\multicolumn{1}{|l|}{$\#$ of IXP address prefixes}  & 312 & 343 \\ \hline
			\multicolumn{1}{|l|}{$\#$ of IXP membership triplets}  & 12,323 & 3,580 \\ \hline
			\hline
			\multicolumn{1}{|l|}{$\#$ of IXPs with membership data} & 448 (88$\%$) & 343 (74$\%$) \\ \hline
			\multicolumn{1}{|l|}{\% of IXPs in top-50 with membership data} & 100$\%$ & 62$\%$ \\ \hline 
			\hline
			\multicolumn{1}{|l|}{$\#$ of IXPs with IP prefix data} & 272 (53$\%$)& 299 (64$\%$) \\ \hline 
			\multicolumn{1}{|l|}{\% of IXPs in top-50 with IP prefix data} & 92$\%$ & 96$\%$ \\ \hline 
		\end{tabular}}
		\caption{Various statistics about the PDB and PCH IXP datasets.\label{tab:pchpdb}}
	\end{table}
	
We then discuss the hit rate of the rules in Tables~\ref{tab:cases1} and~\ref{tab:cases2} in our \texttt{traIXroute} probes to shed more light onto the methodology. The strong evidence Rules 1.1 to 1.3 collectively account for 76.86\% of the detected IXPs, which shows that in most cases we can detect IXPs, while satisfying multiple criteria: \first we observe an exact IP address of a BGP router on the IXP subnet; and \second we find that both ASes are members of the candidate IXP. Rule 1.1 is by far the most frequent as it matches 65.57\% of the detected IXPs. This indicates that the available datasets from PDB and PCH about exact IXP router addresses are rich enough to match most IXP addresses observed in traceroute measurements.
	
Rules 1.4 to 1.7 collectively account for 19.02\% of the matches. These rules rely on weaker evidence. The Rules 1.4 and 1.5, in particular, which rely on IXP prefixes match 13.25\% of the cases. We observe that IXP prefixes add a moderate amount of weak evidence matches compared to the IXP membership data. 
	
Rule 2 hits in 1.36\% of the detected IXPs. This illustrates that in a few cases, the IXP fabric maybe crossed twice. This points to inefficient routing due to the BGP path selection process that relies on AS-level paths and ignores layer-2 topologies. In this case, the layer-2 IXP fabric is likely crossed back and forth, consuming resources.  
	
Besides, we explored a number of other rules, which we do not show in Tables~\ref{tab:cases1} and~\ref{tab:cases2} because they matched in less than 1\% of the cases. From these rules, we confirmed (as expected) that the IXP link is almost always before the observed IXP address. This is because routers typically reply with the IP address of the inbound interface. In just 0.71\% of the cases we observed the IP address, which matched an IXP triplet, to belong to the same AS with the preceding IP address.  Another interesting observation is that when an IP address matches an IXP prefix, but not an IXP triplet, then in only 2.98\% of the matches both of the adjacent ASes are members of the IXP. In contrast, the corresponding number for matched IXP triplets is 81.79\%. This supports further the point that triplets help to detect IXPs more reliably than IXP prefixes.
	
	\begin{table}[t]
		\centering
		\resizebox{220pt}{18pt}{
			\begin{tabular}{|c|c|c|}
				\cline{1-3}
				\textbf{Statistics}  & \textbf{PDB} & \textbf{PCH} \\ \hline \hline 
				%			\multicolumn{1}{|l|}{$\%$ Common tuples \textit{IXP}-\textit{AS member}) with BGP} & 37.8$\%$ & 85.8$\%$\\ \hline 
				\multicolumn{1}{|l|}{\# of (\textit{IXP}-\textit{AS}) tuples in intersection with BGP} & 4,655  & 3,073 \\ \hline 
				\multicolumn{1}{|l|}{$\%$ of tuples (\textit{IXP}-\textit{AS}) with consistent IP addresses} & 93.4$\%$& 92.1$\%$\\ \hline 
			\end{tabular}
		}
		\caption{Consistency of IXP router IP addresses in PDB and PCH with data from 87 BGP Route Collectors located at IXPs}
		\label{tab:validation}
		
	\end{table}
	
\subsection{Data Accuracy and Validation}\label{sec:val} 
The data in PDB are primarily self-reported by IXP and ISP operators, while the data in PCH are based primarily on live BGP Route Collectors that PCH operates in multiple IXP sites, where it is an IXP member and peers with other ASes. The PDB data are often used by network operators for checking and configuring their routers. A recent study~\cite{peeringDB-accuracy} showed that 99\% of the valid (i.e., that conform to the correct format) IP addresses reported in PDB matched the IP addresses used by BGP routers, based on a sample submitted by network operators for 256 routers. We validate further the accuracy of the used PDB and PCH IXP membership data based on BGP feeds from the Route Collectors of PCH.
	
We parse BGP routing table dumps downloaded on January, the 31st 2015 from 87 Route Collectors operated by PCH. Route Collectors on IXPs peer with members of the IXP to provide a live view of their routing announcements. They are therefore an excellent reference for validation because their attributes, e.g. IP addresses and AS numbers, are used in live BGP sessions. For each routing table entry, we extract the next hop IP address and the first AS in the AS path. We then compare the extracted data with the corresponding information from PDB and PCH. We find that 93.4\% of the 4,655 \{\textit{IXP}-\textit{AS}\} tuples, which are common between PDB and BGP, have consistent IP addresses. Accordingly, 92.1\% of the 3,073 \{\textit{IXP}-\textit{AS}\} tuples, which are common between PCH and BGP, have consistent IP addresses. This data is summarized in Table~\ref{tab:validation}. This high degree of consistency shows that triplets \{\textit{IP address} $\longrightarrow$ \textit{IXP}, \textit{AS}\} from PDB and PCH are a valid source of information for detecting IXPs in traceroute paths. The inconsistent part could result from stale or incomplete information in PDB and PCH. Triplets with stale IP addresses will not help, but will not also introduce problems in detecting IXPs with our methodology. Finally, we note that although the triplets we exploit have a reasonable level of accuracy, their completeness is hard to assess. This is a limitation for our work. However, our analysis is encouraging because we find 12,716 triplets for 460 IXPs after merging the PDB and PCH data.
	
Finally, as an extra validation step we cross-checked the Routeviews Prefix to AS mappings Dataset from CAIDA with the IP to ASN mapping service of Team Cymru~\cite{team_cymru} and found that the two mappings were fully consistent. 

\section{Use Case: IXPs in Traceroute Paths}\label{usecase}
Having evaluated and validated our approach, we next do a preliminary analysis of what we can learn about IXPs using an IXP-informed traceroute tool.  We use traceroute paths collected from CAIDA's Ark measurement infrastructure\cite{traceroutes}, which at the time of our experiments had 107 monitors distributed around the globe (split into three teams of similar size). The monitors rely on the scamper tool\cite{luckie2010scamper} configured with the Paris traceroute technique\cite{augustin2007multipath} to mitigate artifacts due to load balancing. We use one full cycle of measurements collected on January, the 20th 2015, which includes an ICMP-paris probe to each globally routed /24 block. Each probe is  assigned to a team. We process the output of scamper with \texttt{traIXroute} to detect IXPs. We repeat our experiments with data from the three teams to check for the consistency of our results across different vantage points. In addition, we process the collected paths to remove probes without any reply or with loops. The number of probes after pre-processing dropped from to 31.8 million to 27.8 million probes.

In Table~\ref{tab:traceroute-stats} we first report the fraction of traceroute paths which go through an IXP. The monitors are located in a mix of academic and corporate institutions~\cite{ark-monitors}. We first observe that the fraction of paths that cross an IXP is 17.44\%, 17.65\% and 23.64\% in the three teams. We observe a slightly larger fraction in the 3rd team, because one of the monitors in this team is located in an IXP (AMS-IX). Despite this, our results are mostly consistent across the three teams: Approximately one out of five paths in our datasets go through an IXP. Furthermore, in paths that go through an IXP we observe 1 to 1.05 IXPs per path. This is interesting because it confirms the expectation based on the valley-free model~\cite{NoGlobalCoor} that up to one peering link, and therefore one IXP\footnote{IXPs links are typically used for settlement-free peering relationships.}, is crossed in an end-to-end path. Even if BGP allows much more complex policies and the Internet IXP ecosystem evolves continuously, Internet paths in our measurements largely conform to the well-known valley-free model. Furthermore, we observe that paths cross on average between 14.06 to 14.77 hops, and the IXP hop is located near the middle, i.e., on average between hop 5.4 and 6.68 for the different teams. For completeness, we also compute the number of ASes the paths cross, which ranges between 4.17 and 4.48 ASes. 

\begin{table}[t]	
	\centering
	\resizebox{190pt}{30pt}{
		\small \begin{tabular}{|c|c|c|c|}
			\cline{1-4}
			\textbf{Statistics}  & \textbf{Team 1} & \textbf{Team 2} & \textbf{Team 3} \\ \hline \hline 
			$\%$Paths with IXPs & 17.65$\%$ & 17.44$\%$ & 23.64$\%$  \\ \hline
			$Avg.$ \# of IXPs per IXP~path & 1.02& 1& 1.05 \\ \hline
			$Avg.$ \# of hops per path & 14.77& 14.37& 14.06\\ \hline
			$Avg.$ IXP hop & 6.68& 6.35& 5.40 \\ \hline
			$Avg.$ \# of ASes per path  & 4.48& 4.17& 4.33 \\ \hline
			%%		$\%$Paths with IXP IPs  & 19.74$\%$ & 19.58$\%$ & 24.85$\%$\\ \hline
		\end{tabular}}
		\caption{Statistics about IXPs in 27.85 million probed traceroute paths. The results are grouped into teams to show the consistency of the computed statistics across vantage points. \label{tab:traceroute-stats}}
	\end{table}
	
{\bf Top IXPs in terms of paths}. We next analyze which IXPs attract most paths and how the number of paths an IXP attracts compares with the number of their member ASes. In Table~\ref{tab:rank} we show the top-10 IXPs in terms of paths, the min and max numbers of paths over the three teams, and number of their members. We first observe that the top-3 IXPs, namely AMS-IX, LINX, and DE-CIX, are the same both in terms of paths and  members. These IXPs are located in Europe; 5 of the following IXPs are located in the US and 4 of these are run by Equinix, i.e., the largest IXP corporation in the US. Finally, one IXP in South America and one more European close the top-10. We note that in Table~\ref{tab:rank} the 570K paths that cross the AMS-IX, is an outlier due to a single ark monitor located in AMS-IX. Despite this, the ranking does not change if we only consider the other teams of monitors.
	
\begin{figure}[t]
			\centering
			\resizebox{250pt}{150pt}{
				\includegraphics{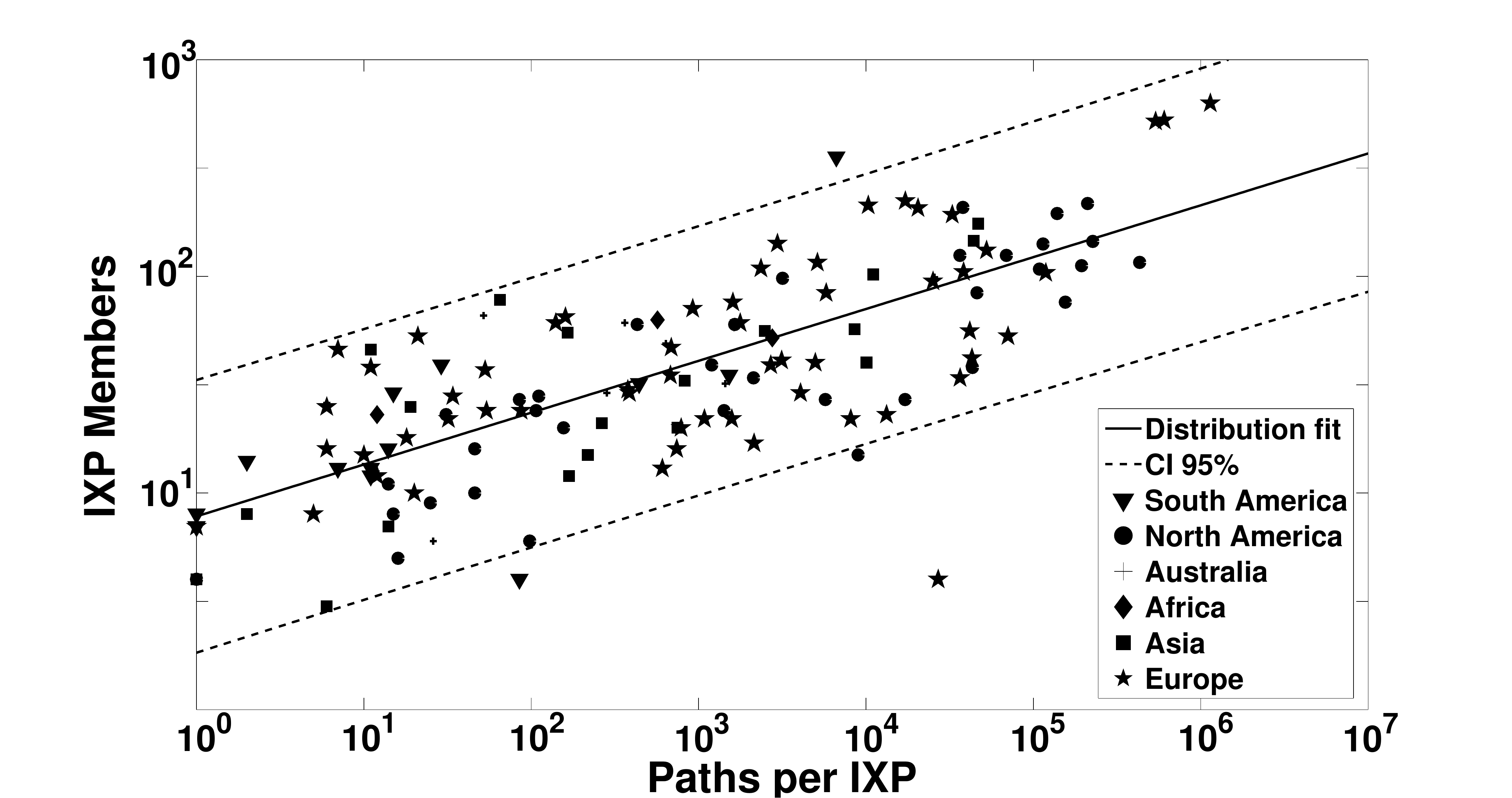}
			}
			%\hspace{10.5cm}
			\caption{Scatterplot of number of AS members vs. number of paths per IXP along with fitted line and 95\% confidence intervals (CI). IXPs are grouped by continent. The correlation is 0.8.}
			\label{fig:correlation}
		\end{figure}

Besides, below the top-3 IXPs we observe significant variance between the number of IXP members and the number of IXP paths. Figure~\ref{fig:correlation} illustrates how the number of IXP members correlates with the number of paths. The overall correlation coefficient is 0.8. We observe that the top-3 IXPs are close to the 95-percentile confidence intervals, which means that compared to the average they have more members than paths. In contrast, many US IXPs have more paths than their number of members indicates. Notably, Equinix Palo Alto is in the 4th position with a small difference in terms of paths from DE-CIX, although the latter has 520 members and the former only 116.

	\begin{table}[t]
		\centering
		\resizebox{230pt}{55pt}{
			\begin{tabular}{|l|c|c|}
				\cline{1-3}
				\textbf{IXP Name} &  \shortstack{\textbf{Min-max \# of paths} \\ \textbf{over teams}}& \textbf{$\#$ of member ASes} \\ \hline \hline 
				1. AMS-IX& 277K - 570K  & 630 \\ \hline 
				2. LINX& 182K - 234K & 526 \\ \hline
				3. DE-CIX Frankfurt& 133K - 215K  & 520 \\ \hline
				4. Equinix Palo Alto& 119K - 134K & 116\\ \hline
				5. Equinix Chicago& 73K - 80K  & 145 \\ \hline
				6. Equinix Ashburn& 43K - 91K & 217 \\ \hline
				7. NAP of The Americas& 45K - 90K  & 112 \\ \hline
				8. Equinix Los Angeles& 37K - 60K & 76 \\ \hline
				9. CoreSite - California& 30K - 58K  & 195 \\ \hline
				10. Netnod Stockholm& 33K - 44K & 104 \\ \hline
			\end{tabular}}
			\caption{Top IXPs sorted by the number of paths that cross them. For each IXP, we show the minimum and maximum number of paths that cross it over the three probing teams; and the number of AS members.	\label{tab:rank}}
		\end{table}
		
\section{Conclusions}\label{conclusion}
Internet users, network operators, and researchers would benefit if they were able to know from which IXPs packets go through. To help towards this goal, in this paper we introduce a tool that extends the commonly used \texttt{traceroute} with techniques to detect IXPs. Our techniques rely on data about the exact IP addresses of BGP router interfaces connected to the IXP subnet, \ie triplets \{\textit{IP address} $\longrightarrow$ \textit{IXP}, \textit{AS}\}, extracted from the \textsc{PeeringDB} and the \textsc{Packet Clearing House}. This data has not been previously explored for identifying IXPs. We show that they are both rich, \ie we find 12,716 triplets for 460 IXPs, and accurate, \ie our validation shows 92-93\% accuracy. We also incorporate in our heuristics an IXP membership check for the adjacent ASes to have stronger evidence that an IXP was crossed. To demonstrate the utility of \texttt{traIXroute}, we use it to show that approximately one out of five paths cross an IXP in our data. In addition, in most cases, we observe not more than one IXP per path, which is located near the middle. Furthermore, we observe that although the top IXPs both in terms of paths and members are located in Europe, US IXPs attract many more paths than their number of members indicates. In the future, we plan to investigate how \texttt{traIXroute} could help Internet users to have more control over their paths.

\section*{Acknowledgements}\label{sec:acknowledgments}
This work has been funded by the European Research Council Grant Agreement no. 338402. We would like to thank Pavlos Sermpezis, Laurent Vanbever, Michalis Bamiedakis and the anonymous reviewers for their helpful comments.

\bibliographystyle{splncs03}
\bibliography{PAM16_85-nomikos}

\end{document}